\documentclass[12pt,a4paper]{article}
\usepackage[dvips]{graphics}
\usepackage{axodraw}
\def\slash#1{\ooalign{\hfil/\hfil\crcr$#1$}}

\makeatletter
\def\dash{-}
\def\@citex[#1]#2{%
\if@filesw\immediate\write\@auxout{\string\citation{#2}}\fi
  \def\@citea{}\@cite{\@for\@citeb:=#2\do
     {\ifx\dash\@citeb{--}\def\@citea{}\else
      \@citea\def\@citea{,\penalty\@m\ }\@ifundefined
       {b@\@citeb}{{\bf ?}\@warning
       {Citation `\@citeb' on page \thepage \space undefined}}\fi
\hbox{\csname b@\@citeb\endcsname}}}{#1}}
\newbox\tempboxa
\newdimen\captionboxsubcount

\newdimen\captionboxsub
\captionboxsub=\hsize \advance\captionboxsub by -\captionboxsubcount
\advance\captionboxsub by -\captionboxsubcount
\long\def\@makecaption#1#2{
 \setbox\@tempboxa\hbox{#1: #2}
 \ifdim \wd\@tempboxa >\captionboxsub
\rightskip=\captionboxsubcount \leftskip=\captionboxsubcount #1: #2
\else \hbox to\hsize{\hfil\box\@tempboxa\hfil}
 \fi}
\makeatother
\begin{document}
%\setlength{\baselineskip}{16pt}

%%%%%%%%%%% title page %%%%%%%%%%%%%

\thispagestyle{empty}

\rightline{DPNU-01-02}
\rightline{OCHA-PP-173}
\rightline{IPPP/01/29}
\rightline{DCPT/01/58}

\vspace{1.5cm}

\centerline{\Large\bf  $B \to K \eta^{(\prime)}$ decay in perturbative QCD}

 \vskip1.5truecm
\renewcommand{\thefootnote}{\fnsymbol{footnote}}
\centerline{\bf E. Kou$\ ^{1,2,3}$
\footnote{ e-mail address: Emi.Kou@durham.ac.uk} 
and A. I. Sanda$\ ^1$
\footnote{ e-mail address: sanda@eken.phys.nagoya-u.ac.jp}}
\bigskip
\centerline{\small\sl 1) Physics Department, Nagoya University, Nagoya 464-8602,
Japan}
\centerline{\small\sl 2) Department of Physics, Ochanomizu University, Tokyo 112-8610,
Japan}
\centerline{\small\sl 3) Institute of Particle Physics Phenomenology, 
University of Durham, Durham, DH1 3LE, U.K.}
 \vskip0.5truecm

\vspace*{1truecm}
\bigskip
\centerline{\large\bf Abstract }

We compute $B \to K \eta^{(\prime)}$ branching ratio
using perturbative QCD approach.
We show that a triangular relation among amplitudes for
$B^0 \to K^0\pi^0$, $B^0 \to K^0 \eta$, $B^0 \to K^0 \eta^{\prime}$
receives large corrections from $SU(3)$ breaking effects.
If experimental value will come closer to the lower limit of 
the present BELLE data there will be a possibility to understand the large 
branching ratio of $B^0 \to K^0\eta^{\prime}$. Otherwise, 
we perhaps need to modify our understanding of $\eta^{\prime}$
meson, for example, inclusion of a possible admixture of gluonium state.

%%% end title page %%%%%%%%%%%%%

\newpage
\renewcommand{\baselinestretch}{2.0}

\pagenumbering{arabic}

%%%%%%%%%%%%%%% INTRODUCTION %%%%%%%%%%%%%%%%%%%%%%%%%%%%5
Three years has passed since CLEO announced an unexpectedly large
branching ratio for $B \to K \eta^{\prime}$ decays ~\cite{CLEO}:
\begin{equation}
Br(\bar{B^0}\to \bar{K^0} \eta^{\prime})= (89^{+18}_{-16}\pm 9)\times 10^{-6}
\label{eq:1}
\end{equation}
BELLE also reported their results in $BCP4$ conference, ~\cite{BELLE}:
\begin{equation}
Br(\bar{B^0}\to \bar{K^0} \eta^{\prime})= (64^{+25+10}_{-20-11})\times 10^{-6}
\label{eq:2-3}
\end{equation}
Various theoretical suggestions have been made to understand
the large branching ratio.
While new physics contributions were discussed ~\cite{HOU,KAGAN}
we feel that better understanding of the standard model calculation of the
branching ratio is necessary.
In Ref. ~\cite{PACV}, it was shown that there is
a possible choice of theoretical parameters involving
form factors, CKM parameters, nonfactorizable contribution
and decay constants of $\eta - \eta^{\prime}$ system
which gives a branching ratio consistent with the experimental data.
A $SU(3)$ relation which is independent of most of the above mentioned
uncertainties
has been derived ~\cite{LIP,ROS,LIP2}:
\begin{equation}
-3\sqrt{2}A(B^0\to K^0 \pi^0)+4\sqrt{3}A(B^0\to K^0 \eta)=
\sqrt{6}A(B^0\to K^0 \eta^{\prime}). \label{eq:3}
\end{equation}
Using the CLEO measurement ~\cite{CLEO2}, 
$Br(\bar{B^0}\to \bar{K^0} \pi^0)=
(14.6^{+5.9+2.4}_{-5.1-3.3})\times 10^{-6}$, and theoretical
expectation
that $|A(B^0\to K^0\eta)|$ is small compared to the other two amplitudes,
the observed value for $Br(\bar{B^0} \to \bar{K^0} \eta^{\prime})$
in Eq. (\ref{eq:1}) seems to be too large. This relation also excludes
explanations which make $Br(\bar{B^0} \to \bar{K^0}\pi^0)$ 
increase simultaneously with  $Br(\bar{B^0} \to \bar{K^0}\eta^{\prime})$, 
for instance, invoking large Wilson coefficients
with new physics effects, or increasing the input parameters like form factors,
the CKM parameters, etc.

In this letter, we perform calculation of the branching
ratio by using  perturbative QCD (pQCD) approach and examine
the $B \to K \eta^{\prime}$ problem.
The $SU(3)$ breaking effect is included through the decay
constants and the wave function and as a result,
Eq. (\ref{eq:3}) is  modified.
We also give a theoretical estimate of
color suppressed penguin contributions which is one of the candidate
mechanism to enhance $Br(B^0 \to K^0\eta^{(\prime)})$, but not
$Br(\bar{B^0} \to \bar{K^0}\pi^0)$.

In the 80's, $\eta^{\prime}$ gluonic admixture was
examined in Ref. ~\cite{ETA}. Recently, there have been some progress in understanding the
$\eta-\eta^{\prime}$ system. We use the $\eta-\eta^{\prime}$ mixing angle and
the definition of the decay constant in $\eta -\eta^{\prime}$ system 
which include recent improvements.
A simple description of $\eta-\eta^{\prime}$ states was
introduced in the literature,
\begin{eqnarray}
        |\eta>&=&X_{\eta}|\frac{u\bar{u}+d\bar{d}}{\sqrt{2}}>+
        Y_{\eta}|s\bar{s}> \\
  |\eta^{\prime}>&=&X_{\eta^{\prime}}|\frac{u\bar{u}+d\bar{d}}{\sqrt{2}}>+
  Y_{\eta^{\prime}}|s\bar{s}> +Z_{\eta^{\prime}}|gluonium>,
\end{eqnarray}
where $X_{\eta^{(\prime)}}$, $Y_{\eta^{(\prime)}}$ and $Z_{\eta^{\prime}}$
parameters represent the ratios of $u\bar{u}+d\bar{d}$, $s\bar{s}$ and
gluonium component of $\eta^{(\prime)}$, respectively.
This work was updated by one of the authors ~\cite{KOU}.
In this work, the gluonium content of $\eta^{\prime}$ is
reanalyzed using all available radiative light meson decays and
a result
\begin{equation}
\frac{Z_{\eta^{\prime}}}{X_{\eta^{\prime}}+Y_{\eta^{\prime}}+
Z_{\eta^{\prime}}}\leq 0.26
\end{equation}
is obtained, which indicates that 26\% of gluonic admixture in
$\eta^{\prime}$ is still possible.

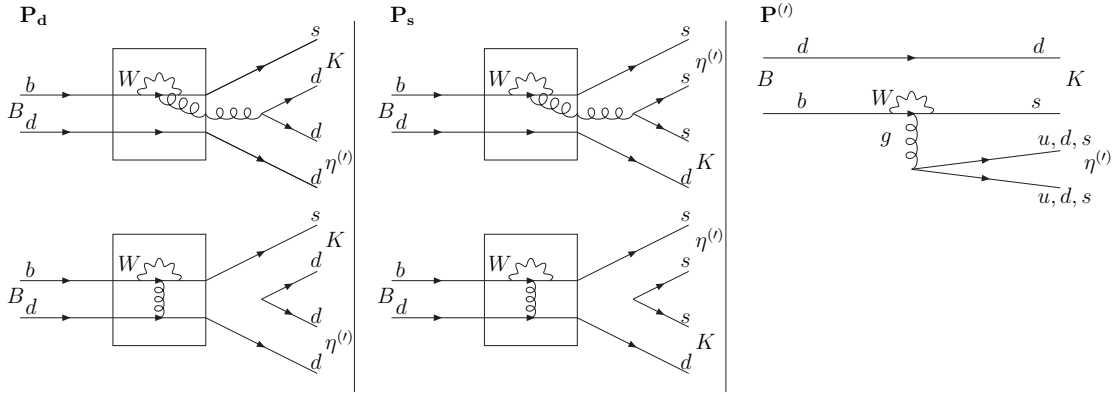
\begin{figure}[h]
 \begin{center}
\begin{center} \scalebox{0.7}{\begin{picture}(600,200)(0,-100)
\put(20,100){$\mathbf{P_d}$}
\ArrowLine(20,60)(70,60)\ArrowLine(20,40)(70,40)
\BBox(70,25)(120,85)
\ArrowLine(70,60)(120,60)\ArrowLine(70,40)(120,40)
\PhotonArc(95,60)(10,0,180){2}{4.5}\Gluon(95,60)(120,50){-3.5}{3.5}
\ArrowLine(120,60)(180,90)\ArrowLine(120,40)(180,10)
\Gluon(120,50)(150,50){-3}{3}
\ArrowLine(150,50)(180,65)\ArrowLine(150,50)(180,35)
\put(23,62){$b$}\put(23,42){$d$}
\put(177,92){$s$}\put(177,67){$d$}\put(177,37){$d$}\put(177,12){$d$}
\put(13,48){$B$}\put(185,75){$K$}\put(185,20){$\eta^{(\prime)}$}
\put(73,65){$W$}
\ArrowLine(20,-60)(70,-60)\ArrowLine(20,-40)(70,-40)
\BBox(70,-15)(120,-75)
\ArrowLine(70,-60)(120,-60)\ArrowLine(70,-40)(120,-40)
\PhotonArc(95,-40)(10,0,180){2}{4.5}\Gluon(95,-60)(95,-40){-2.5}{3.5}
\ArrowLine(120,-60)(180,-90)\ArrowLine(120,-40)(180,-10)
\ArrowLine(150,-50)(180,-65)\ArrowLine(150,-50)(180,-35)
\put(23,-38){$b$}\put(23,-58){$d$}
\put(177,-63){$d$}\put(177,-88){$d$}\put(177,-8){$s$}\put(177,-33){$d$}
\put(13,-52){$B$}\put(185,-77){$\eta^{(\prime)}$}\put(185,-22){$K$}
\put(73,-35){$W$}
\put(220,100){$\mathbf{P_s}$}
\ArrowLine(220,60)(270,60)\ArrowLine(220,40)(270,40)
\BBox(270,25)(320,85)
\ArrowLine(270,60)(320,60)\ArrowLine(270,40)(320,40)
\PhotonArc(295,60)(10,0,180){2}{4.5}\Gluon(295,60)(320,50){-3.5}{3.5}
\ArrowLine(120,60)(180,90)\ArrowLine(120,40)(180,10)
\ArrowLine(320,60)(380,90)\ArrowLine(320,40)(380,10)
\Gluon(320,50)(350,50){-3}{3}
\ArrowLine(350,50)(380,65)\ArrowLine(350,50)(380,35)
\put(223,62){$b$}\put(223,42){$d$}
\put(377,92){$s$}\put(377,67){$s$}\put(377,37){$s$}\put(377,12){$d$}
\put(213,48){$B$}\put(385,75){$\eta^{(\prime)}$}\put(385,20){$K$}
\put(273,65){$W$}
\ArrowLine(220,-60)(270,-60)\ArrowLine(220,-40)(270,-40)
\BBox(270,-15)(320,-75)
\ArrowLine(270,-60)(320,-60)\ArrowLine(270,-40)(320,-40)
\PhotonArc(295,-40)(10,0,180){2}{4.5}\Gluon(295,-60)(295,-40){-2.5}{3.5}
\ArrowLine(320,-60)(380,-90)\ArrowLine(320,-40)(380,-10)
\ArrowLine(350,-50)(380,-65)\ArrowLine(350,-50)(380,-35)
\put(223,-38){$b$}\put(223,-58){$d$}
\put(377,-63){$s$}\put(377,-88){$d$}\put(377,-8){$s$}\put(377,-33){$s$}
\put(213,-52){$B$}\put(385,-77){$K$}\put(385,-22){$\eta^{(\prime)}$}
\put(273,-35){$W$}
\put(420,100){$\mathbf{P^{(\prime)}}$}
\ArrowLine(420,50)(580,50)
\PhotonArc(500,50)(10,0,180){2}{4.5}
\Gluon(500,50)(500,20){3}{3}
\ArrowLine(500,20)(580,30)
\ArrowLine(500,20)(580,10)
\ArrowLine(420,80)(580,80)
\put(440,53)
{$b$\ \ \ \ \ \ \ \ \ \ \ \ \ \ \ \ \ \ \ \ \ \ \ \ \ \ \ \ \ \ \ $s$}
\put(440,83)
{$d$\ \ \ \ \ \ \ \ \ \ \ \ \ \ \ \ \ \ \ \ \ \ \ \ \ \ \ \ \ \ \ $d$}
\put(570,33){$u,d,s$} \put(570,2){$u,d,s$}
\put(480,55){$W$} \put(485,35){$g$}
\put(418,65){$B$} \put(585,65){$K$} \put(595,20){$\eta^{(\prime)}$}
\Line(200,100)(200,-100) \Line(400,100)(400,-100)
\end{picture}} \end{center}
 \end{center}
\vspace{-1cm}%)
 \caption[fig1]
{\vspace{-0.5cm}
Diagrams for $bsd\bar{d}$ penguin, $P_d$, $bss\bar{s}$ penguin, $P_s$ and
$SU(3)$ singlet penguin, $P^{\prime}$ contributions}
 \label{fig:1-2}
\end{figure}

Ignoring the small tree contribution, we can write the amplitudes
of $B^0 \to K^0 \pi^0$ and $B^0 \to K^0 \eta^{(\prime)}$ decays as:
\begin{eqnarray}
A(B^0 \to K^0 \pi)&=&-1/\sqrt{2}P_d \label{eq:6-2}\\
A(B^0 \to K^0 \eta)&=&
  X_{\eta}/\sqrt{2}P_d+Y_{\eta}P_s+P \label{eq:7-2}\\
A(B^0 \to K^0 \eta^{\prime})&=&
  X_{\eta^{\prime}}/\sqrt{2}P_d+Y_{\eta^{\prime}}P_s+P^{\prime},
\label{eq:8-2}
\end{eqnarray}
where $P_{d(s)}$ includes color allowed $bsd\bar{d}(s\bar{s})$ penguin
and annihilation penguin contributions and $P^{(\prime)}$
is $SU(3)$ singlet contribution.
We depict the corresponding diagrams
in Fig.\ref{fig:1-2}.

Here we introduce a parameter $r$ which represents the $SU(3)$
breaking effect:
\begin{equation}
P_s=rP_d
\label{eq:11-7}
\end{equation}
and a parameter $s^{(\prime)}$ which represents the ratio between
$P^{(\prime)}$ and $P_s$:
\begin{equation}
P=sP_s, \ \ \ \ \ \ \ \ \ \ P^{\prime}=s'P_s. \label{eq:11-11}
\end{equation}
Using these parameters, Eqs. (\ref{eq:6-2}-\ref{eq:8-2}) lead to
$SU(3)$ relation in general form
\begin{eqnarray}
A(B^0\to K^0\eta)
&=&-(X_\eta+r\sqrt{2}Y_\eta+r\sqrt{2}s)
A(B^0\to K^0\pi^0) \label{eq:12-5} \\
A(B^0\to K^0\eta^{\prime})
&=&-(X_{\eta'}+r\sqrt{2}Y_{\eta'}+r\sqrt{2}s')
A(B^0\to K^0\pi^0). \label{eq:13-5}
\end{eqnarray}

Now, let us examine what it takes to obtain
Eq. (\ref{eq:3}). The following assumptions must be applied:
\begin{itemize}
\item
$\eta^{\prime}$ does not have the gluonic content
($Z_{\eta^{\prime}}=0$) so that  $X_{\eta (\eta^{\prime})}$
and  $Y_{\eta (\eta^{\prime})}$ are related
to the pseudoscalar mixing angles as
\begin{equation}
        \left(\begin{array}{@{\,}c@{\,}}
        \eta \\ \eta^{\prime} \end{array} \right)
   = \left(\begin{array}{@{\,}cccc@{\,}}
                                \cos{\alpha_p} & -\sin{\alpha_p} \\
                                \sin{\alpha_p} & \cos{\alpha_p} \\
                         \end{array} \right)
           \left(\begin{array}{@{\,}c@{\,}}
           \frac{u\bar{u}+d\bar{d}}{\sqrt{2}} \\
            s\bar{s} \end{array} \right).
\label{defalph}
\end{equation}
For convenience, we also display $\eta$ and $\eta^{\prime}$ states 
in terms of singlet state $\eta_1$ and octet state $\eta_8$ as
\begin{equation}
        \left(\begin{array}{@{\,}c@{\,}}
        \eta \\ \eta^{\prime} \end{array} \right)
   = \left(\begin{array}{@{\,}cccc@{\,}}
                                \cos{\theta_p} & -\sin{\theta_p} \\
                                \sin{\theta_p} & \cos{\theta_p} \\
                         \end{array} \right)
           \left(\begin{array}{@{\,}c@{\,}}
           \eta_8 \\ \eta_1 \end{array} \right),  \nonumber
\label{eq:15-100}
\end{equation}
where $\eta_8=(u\bar{u}+d\bar{d}-2s\bar{s})/\sqrt{6}$ and
$\eta_1=(u\bar{u}+d\bar{d}+s\bar{s})/\sqrt{3}$.
$\alpha_p$ can be written in terms of
the pseudoscalar mixing angle $\theta_p$
as $\alpha_p=\theta_p-\theta_I+\frac{\pi}{2}$ with the ideal mixing
$\theta_I=\arctan(1/\sqrt{2})$. The allowed range of values for $\theta_p$ is
$-20^{\circ}$ to $-10^{\circ}$. To obtain Eq. (\ref{eq:3}), the angle $\alpha_p$ is fixed at
$\cos\alpha_p=\sqrt{2}/\sqrt{3}$ and $\sin\alpha_p=1/\sqrt{3}$,
which corresponds to $\theta_p\approx -19.4^{\circ}$.
We should remind the reader that we are taking a value of $\theta_p$ 
which is at the edge of the allowed region. 
While we stretched the error bar
to allow the region $\theta_p\sim -20^\circ$,  the experimental data for
$\omega\to\eta\gamma$ decay and $\eta\to\gamma\gamma$ decay
disfavor this region. The best fit range is:
 $-17^{\circ}$ to $-10^{\circ}$ in ~\cite{KOU}.
\item
$SU(3)$ symmetry is exact so that $P_d=P_s$,
{\it i.e.} $r=1$.

\item
The ratio of the $SU(3)$ flavor singlet contribution
to $B \to K\eta$ and $B \to K\eta^{\prime}$ is written as
$s'/s=
-\cos\theta_p/\sin\theta_p$, which is extracted from the ratio
of the $SU(3)$ singlet component of $\eta$ and $\eta^{\prime}$
states in Eq. (\ref{eq:15-100}). 
It can be easily seen below that this is valid only if we set $f_K=f_\pi$ 
and ignore the electric penguin correction factor ($\xi=1$). 
\end{itemize}
We would like to point out that whether 
the experimental value Eq.(\ref{eq:1}) is 
inconsistent with Eq. (\ref{eq:3}) 
depends crucially on the assumptions above. 
Looking at the $SU(3)$ relation in general form in Eq. (\ref{eq:13-5}), 
we see that the amplitude of $B^0 \to K^0 \eta^{\prime}$ can be enhanced 
by large $r$ and $s'$. 
And in fact, relaxing above assumptions, 
we can easily have $Br(B^0 \to K^0 \eta^{\prime})$ 
consistent with data. 
For example, if we choose $r=1.1$ and $\theta_p=-10^{\circ}$, 
keep the relation $s'/s=-\cos\theta_p/\sin\theta_p$ and take 
$Br(B^0 \to K^0\pi^0)=15\times 10^{-6}$ 
and $Br(B^0 \to K^0 \eta)=0$, Eqs. (\ref{eq:12-5}) and (\ref{eq:13-5}) give  
\begin{equation}
Br(B^0 \to K^0 \eta^{\prime})=84\times 10^{-6}.
\label{eq:1000}
\end{equation}
We insist that $SU(3)$ breaking effects for $r$, $s$ and $s'$ 
must be studied before 
we conclude that experimental data is too large - and that we need 
new physics to explain the observations.
The large value in Eq. (\ref{eq:1000}) is due to the fact that
constraints:
$Br(B^0 \to K^0 \eta)=0$ and $s'/s=-\cos\theta_p/\sin\theta_p$ 
leads to large $s'$. 

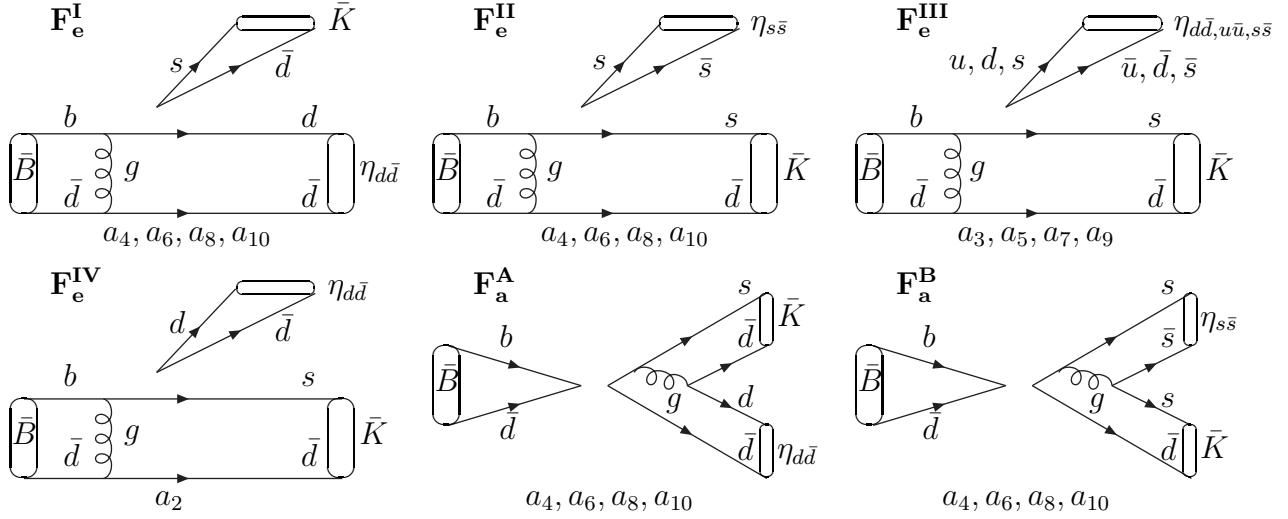
\begin{figure}[ht]
 \begin{center}
\vspace{0.5cm}
\begin{picture}(500,200)(60,0)
\ArrowLine(30,110)(150,110) \ArrowLine(30,140)(150,140)
\ArrowLine(80,150)(109.8,181.5) \ArrowLine(80,150)(139.8,179.8)
\Gluon(60,140)(60,110){3}{3} \put(68,125){$g$}
\put(30,125.5){\oval(10,30)} \put(150,125.5){\oval(10,30)}
\put(125.3,182.3){\oval(29.2,5)}
\put(25.5,123){$\bar{B}$} \put(157,125){$\eta_{d\bar{d}}$}
\put(145,180){$\bar{K}$} \put(45,143){$b$}
\put(135,143){$d$} \put(45,113){$\bar{d}$} \put(135,113){$\bar{d}$}
\put(85,164){$s$} \put(125,162){$\bar{d}$}

\ArrowLine(190,110)(310,110) \ArrowLine(190,140)(310,140)
\ArrowLine(240,150)(269.8,181.5) \ArrowLine(240,150)(299.7,179.7)
\Gluon(220,140)(220,110){3}{3} \put(228,125){$g$}
\put(190,125.5){\oval(10,30)} \put(310,125.5){\oval(10,30)}
\put(285.3,182.3){\oval(29.4,5)}
\put(185.5,123){$\bar{B}$} \put(317,123){$\bar{K}$}
\put(305,180){$\eta_{s\bar{s}}$}
\put(205,143){$b$} \put(295,143){$s$}
\put(205,113){$\bar{d}$} \put(295,113){$\bar{d}$}
\put(245,165){$s$} \put(285,162){$\bar{s}$}

\ArrowLine(350,110)(470,110) \ArrowLine(350,140)(470,140)
\ArrowLine(400,150)(429.8,181.5) \ArrowLine(400,150)(459.4,179.4)
\Gluon(380,140)(380,110){3}{3} \put(388,125){$g$}
\put(350,125.5){\oval(10,30)} \put(470,125.5){\oval(10,30)}
\put(445.6,182.3){\oval(29.6,5)}
\put(345.5,123){$\bar{B}$} \put(477,123){$\bar{K}$}
\put(465,180){$\eta_{d\bar{d},u\bar{u},s\bar{s}}$}
\put(365,143){$b$} \put(455,143){$s$}
\put(365,113){$\bar{d}$} \put(455,113){$\bar{d}$}
\put(380,165){$u,d,s$} \put(445,162){$\bar{u},\bar{d},\bar{s}$}

\put(60,100){$a_4, a_6, a_8, a_{10}$
\ \ \ \ \ \ \ \ \ \ \ \ \ \ \ \ \ \ \ \ \ \ \ \ \
$a_4, a_6, a_8, a_{10}$\ \ \ \ \ \ \ \ \ \ \ \ \ \ \ \ \ \ \ \ \ \ \ \
$a_3, a_5, a_7, a_9$}

\ArrowLine(30,10)(150,10) \ArrowLine(30,40)(150,40)
\ArrowLine(80,50)(109.8,81.5) \ArrowLine(80,50)(139.8,79.8)
\Gluon(60,40)(60,10){3}{3} \put(68,25){$g$}
\put(30,25.4){\oval(10,30)} \put(150,25.4){\oval(10,30)}
\put(125.3,82.3){\oval(29.2,5)}
\put(25.5,23){$\bar{B}$} \put(157,23){$\bar{K}$}
\put(145,80){$\eta_{d\bar{d}}$}
\put(45,45){$b$} \put(135,45){$s$} \put(45,15){$\bar{d}$}
\put(135,15){$\bar{d}$}
\put(85,65){$d$} \put(125,62){$\bar{d}$}

\ArrowLine(190,60)(240,45) \ArrowLine(190,30)(240,45)
\ArrowLine(250,45)(310,80) \ArrowLine(250,45)(310,10)
\ArrowLine(280,45)(310,60) \ArrowLine(280,45)(310,30)
\Gluon(260,50)(280,45){3}{2} \put(273,37){$g$}
\put(189.8,45.4){\oval(10,30)} \put(311,70.3){\oval(5,19.8)}
\put(311,20.6){\oval(5,19.8)}
\put(185.5,43){$\bar{B}$} \put(315,68){$\bar{K}$}
\put(315,18){$\eta_{d\bar{d}}$}
\put(210,60){$b$} \put(210,25){$\bar{d}$}
\put(300,80){$s$} \put(300,60){$\bar{d}$}
\put(300,37){$d$} \put(300,17){$\bar{d}$}

\ArrowLine(350,60)(400,45) \ArrowLine(350,30)(400,45)
\ArrowLine(410,45)(470,80) \ArrowLine(410,45)(470,10)
\ArrowLine(440,45)(470,60) \ArrowLine(440,45)(470,30)
\Gluon(420,50)(440,45){3}{2} \put(433,37){$g$}
\put(349.8,45.4){\oval(10,30)} \put(471.3,70.3){\oval(5,19.8)}
\put(471,20.6){\oval(5,19.8)}
\put(345.5,43){$\bar{B}$} \put(475,68){$\eta_{s\bar{s}}$}
\put(475,18){$\bar{K}$}
\put(370,60){$b$} \put(370,25){$\bar{d}$}
\put(460,80){$s$} \put(460,60){$\bar{s}$}
\put(460,37){$s$} \put(460,17){$\bar{d}$}

\put(60,0){\ \ \ \ \ $a_2$\ \ \ \ \ \ \ \ \ \ \ \ \ \ \
\ \ \ \ \ \ \ \ \ \ \ \ \ \ \ \ \ \
$a_4, a_6, a_8, a_{10}$ \ \ \ \
\ \ \ \ \ \ \ \ \ \ \ \ \ \ \ \ \ \ \
$a_4, a_6, a_8, a_{10}$}

\put(40,180){$\mathbf{F_e^I}$} \put(200,180){$\mathbf{F_e^{II}}$}
\put(360,180){$\mathbf{F_e^{III}}$} \put(40,80){$\mathbf{F_e^{IV}}$}
\put(200,80){$\mathbf{F_a^A}$} \put(360,80){$\mathbf{F_a^B}$}
\end{picture}
 \end{center}
\vspace{-0.5cm}%)
 \caption[fig5]
{\vspace{-0.5cm}
Contributing diagrams. }
\label{fig:5}
\end{figure}

In our pQCD approach, we 
evaluate these $SU(3)$ breaking parameters as well as the branching 
ratios $Br(\bar{B^0} \to \bar{K^0} \eta^{\prime})$ and 
$Br(\bar{B^0} \to \bar{K^0} \eta)$. 
Now let us explain our calculation for
$\bar{B^0} \to \bar{K^0} \eta^{(\prime)}$ decay amplitude.
(Full calculation will be presented in elsewhere.) 
The pQCD approach is developed to give more
precise theoretical prediction beyond vacuum saturation approximation 
~\cite{LKS,LUY}. In this approach,
the amplitude for $B^0 \to K^0 \eta^{(\prime)}$ is
given as (see also Fig. \ref{fig:5}):
\begin{eqnarray}
\hspace{-0.4cm}
M(\bar{B}^0\to \bar{K}^0\eta^{(\prime)})&=&\frac{M_B^2G_F}{2\sqrt{2}}\large\{
        V_tf_K\frac{X_{\eta^{(\prime)}}}{\sqrt{2}}F_e^{I}+
        V_tf_yY_{\eta^{(\prime)}}F_e^{II} \nonumber \\ &\ &
        +V_t(f_x\frac{X_{\eta^{(\prime)}}}{\sqrt{2}}\xi+
        f_x\frac{X_{\eta^{(\prime)}}}{\sqrt{2}}+
        f_yY_{\eta^{(\prime)}})F_e^{III}
        \nonumber \\ &\ &
        -V_uf_x\frac{X_{\eta^{(\prime)}}}{\sqrt{2}}F_e^{IV}+
        V_tf_B\frac{X_{\eta^{(\prime)}}}{\sqrt{2}}F_a^{A}+
         V_tf_BY_{\eta^{(\prime)}}F_a^{B}\large\}, \label{eq:20}
\end{eqnarray}
where $\xi$ is a correction factor which represents the difference
between electric penguin contributions of
$bsu\bar{u}$ and $bsd\bar{d}(bss\bar{s})$ penguin diagrams and 
$V_q=V_{qs}^*V_{qb}$  $(q=t,u)$. 
The parameter $a_i(\mu)$ in Fig. \ref{fig:5} is defined as 
$a_2(\mu)=C_1(\mu)+C_2(\mu)/N$, $a_i(\mu)=C_i(\mu)+C_{i+1}(\mu)/N$ 
for $i=3,5,7,9$ and $a_i(\mu)=C_i(\mu)+C_{i-1}(\mu)/N$ for $i=4,6,8,10$, 
where $N$ is number of the color and 
$C_i(\mu)$ is Wilson coefficient  
(we use the same definition 
of Wilson coefficient as the one in ~\cite{LKS}). It is worthwhile pointing
out that scale
$\mu$ in these coefficients is related to the loop integration variable
for diagrams shown in Fig. 2. Thus the usual scale dependence problem 
associated with the factorization assumption is absent in the pQCD approach.

\begin{figure}[hb]
 \begin{center}
\scalebox{0.7}{
\begin{picture}(400,100)(0,0)
\ArrowLine(15,10)(185,10) \ArrowLine(15,50)(185,50)
\ArrowLine(90,60)(120,90) \ArrowLine(90,60)(170,90)
\Gluon(130,10)(130,75){5}{6}
\put(15,30.5){\oval(10,40)} \put(185,30.5){\oval(10,40)}
\put(146,93){\oval(50,5)} \put(11,31){$B$}
\ArrowLine(210,70)(290,50) \ArrowLine(210,30)(290,50)
\ArrowLine(300,50)(385,100) \ArrowLine(300,50)(385,0)
\ArrowLine(340,50)(385,75) \ArrowLine(340,50)(385,25)
\put(210,50.3){\oval(10,40)} \put(386.3,87.8){\oval(5,25)}
\put(386,13){\oval(5,25)} \put(205,50){$B$}
\GlueArc(290,0)(70,45,125){5}{8}
\end{picture}}
 \end{center}
\vspace{-1cm}%)
 \caption[fig6]
{\vspace{-0.5cm}
Nonfactorizable contribution}
 \label{fig:6}
\end{figure}
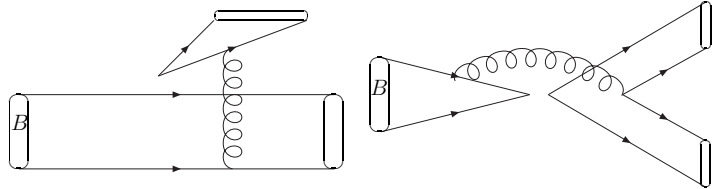

In our notation, $F_e^i$ ($i=I\sim III(IV)$) represents penguin (tree)diagram 
and $F_a^i$ ($i=A,B$) represents annihilation penguin diagram.
The correspondence between Eq. (\ref{eq:8-2}) and (\ref{eq:20})
is as follows:
\begin{eqnarray}
P_d &=& \frac{M_B^2G_F}{2\sqrt{2}}V_t(f_KF_e^I+f_BF_a^A) \\
P_s &=& \frac{M_B^2G_F}{2\sqrt{2}}V_t(f_yF_e^{II}+f_BF_a^B) \\
P^{(\prime)}&=&
        \frac{M_B^2G_F}{2\sqrt{2}}
        V_t(f_x\frac{X_{\eta^{(\prime)}}}{\sqrt{2}}\xi+
        f_x\frac{X_{\eta^{(\prime)}}}{\sqrt{2}}+
        f_yY_{\eta^{(\prime)}})F_e^{III}.
\end{eqnarray}
Note that we also have additional contribution from nonfactorizable diagrams
(see, Fig. \ref{fig:6}), which is not calculated in vacuum saturation
approximation but can be calculable in pQCD approach. However,
we found that nonfactorizable contributions to branching ratios for
$\bar{B^0} \to \bar{K^0} \eta^{\prime}$ and
$\bar{B^0} \to \bar{K^0} \eta$ are less than 10\%.

As is mentioned in the introduction, we use new definitions of decay
constants in the $\eta-\eta^{\prime}$ system -
the decay constants at the non-anomaly limit
~\cite{KF,-,LEU}:
\begin{eqnarray}
if_xp_{\mu} &=&
<0|(u\gamma^{\mu}\gamma_5\bar{u}+d\gamma^{\mu}\gamma_5\bar{d})/\sqrt{2}|
\frac{u\bar{u}+d\bar{d}}{\sqrt{2}}> \\
if_yp_{\mu} &=& <0|s\gamma^{\mu}\gamma_5\bar{s}|s\bar{s}>.
\end{eqnarray}
For the value of $f_x$ and $f_y$, we use the ones that are given
in~\cite{KF} where isospin symmetry is assumed for $f_x$
and $SU(3)$ breaking effect is included for $f_y$:
\begin{equation}
 f_x=f_{\pi}, \ \ \ f_y=\sqrt{2f_K^2-f_{\pi}^2}.\ \ \
\label{eq:7-5}
\end{equation}
These values are translated to the values in the two mixing angle
method, which is often used in vacuum saturation approach as:
\begin{eqnarray}
&f_8=169  \ \mbox{MeV}, \ \ \ \ \ \ \ \ \ \ f_1=151  \ \mbox{MeV}, & \\
&\theta_8=-28.9^{\circ} (-18.9^{\circ}), \ \ \
\theta_1=-10.1^{\circ} (-0.1^{\circ}), & 
\end{eqnarray}
where the pseudoscalar mixing angle $\theta_p$ is taken as
$-20^{\circ}$ ($-10^{\circ}$).
The wave function for $d\bar{d}$ components in
$\eta^{(\prime)}$ and K meson are given as:
\begin{eqnarray}
\Psi_{\eta_{d\bar{d}}}(P,x,\zeta)\equiv
\slash{P}\phi_{\eta_{d\bar{d}}}^A(x)+m_0^{\eta_{d\bar{d}}}
\phi_{\eta_{d\bar{d}}}^P(x)+\zeta m_0^{\eta_{d\bar{d}}}
(\slash{v}\slash{n}-v\cdot n)\phi_{\eta_{d\bar{d}}}^{\sigma \prime}(x)
\\
\Psi_K(P,x,\zeta) \equiv \slash{P}\phi_K^A(x)+m_0^K
\phi_K^P(x)+\zeta m_0^K
(\slash{v}\slash{n}-v\cdot n)\phi_K^{\sigma \prime}(x),
\end{eqnarray}
where $P$ and $x$ are the momentum and the momentum fraction of
$\eta_{d\bar{d}}(K)$, respectively. We assumed here that the wave function 
of $\eta_{d\bar{d}}$ is same as the $\pi$ wave function. 
The parameter $\zeta$ is either $+1$ or $-1$ depending on the assignment 
of the momentum fraction $x$. $\phi_{\eta_{d\bar{d}}(K)}^A$,
$\phi_{\eta_{d\bar{d}}(K)}^P$ and $\phi_{\eta_{d\bar{d}}(K)}^{\sigma}$
represent the axial vector, pseudoscalar and tensor components of
the wave function, respectively, for which we utilize the result
from the Light-Cone sum rule~\cite{BALL} including twist-3 contribution:
\begin{eqnarray}
\phi_{\eta_{d\bar{d}}}^{A}(x)&=&\frac{3}{\sqrt{2N_c}}f_xx(1-x)
[1+a_2^{\eta_{d\bar{d}}}\frac{3}{2}(5(1-2x)^2-1)]+ \nonumber \\
&\ & \ \ \ \ \ \ \ \ \ \ \ \ \ \ \ \ \
a_4^{\eta_{d\bar{d}}}\frac{15}{8}(21(1-2x)^4-14(1-2x)^2+1)]
\nonumber \\
\phi^{P}_{\eta_{d\bar{d}}}(x)&=&\frac{1}{2\sqrt{2N_c}}f_x
[1+(30\eta_3-\frac{5}{2}\rho^2_{\eta_{d\bar{d}}})\frac{1}{2}(3(1-2x)^2-1)+
\nonumber \\
&\ &\ \ \ (-3\eta_3\omega_3-\frac{27}{20}\rho^2_{\eta_{d\bar{d}}}-
\frac{81}{10}\rho^2_{\eta_{d\bar{d}}}a_2^{\eta_{d\bar{d}}})\frac{1}{8}
(35(1-2x)^4-30(1-2x)^2+3)] \nonumber \\
\phi^{\sigma \prime}_{\eta_{d\bar{d}}}(x)
&=&\frac{3}{\sqrt{2N_c}}f_x(1-2x) \nonumber \\
&\ & \ \ \ \ \ [\frac{1}{6}+(5\eta_3-\frac{1}{2}\eta_3\omega_3-
\frac{7}{20}\rho_{\eta_{d\bar{d}}}^2
-\frac{3}{5}\rho^2_{\eta_{d\bar{d}}}a_2^{\eta_{d\bar{d}}})(10x^2-10x+1)] 
\nonumber \\
\phi_K^A(x) &=& \frac{3}{\sqrt{2N_c}}f_Kx(1-x)
[1+3a_1^K(1-2x)+\frac{3}{2}a_2^K(5(1-2x)^2-1)], \nonumber
\end{eqnarray}
where 
\begin{eqnarray}
&a^{\eta_{d\bar{d}}}_2=0.44, \ \ a^{\eta_{d\bar{d}}}_4=0.25, \ \ 
a_1^K=0.20, \ \ a_2^K=0.25, & \nonumber \\ 
&\rho_{\eta_{d\bar{d}}}=\frac{m_{\pi}}{m_0^{\eta_{d\bar{d}}}}, 
\ \ \eta_3=0.015, \ \ \omega_3=-3. & \nonumber 
\end{eqnarray}
We assume that the wave function of $u\bar{u}$ is 
same as the wave function of $d\bar{d}$. 
For the wave function of the $s\bar{s}$ components, we also use the same 
form as $d\bar{d}$ but with $m^{s\bar{s}}_0$ and $f_y$ instead of 
$m^{d\bar{d}}_0$ and $f_x$, respectively. 
The pseudoscalar and tensor components of the $K$ wave function
are obtained by exchanging parameters of the pseudoscalar and
tensor components of $d\bar{d}$ wave function, respectively as follows:
\begin{equation} 
f_x\leftrightarrow f_K, 
\ \ \rho_{\eta_{d\bar{d}}}\leftrightarrow \rho_K=\frac{m_K}{m_0^K} , \ \
a_{2(4)}^{\eta_{d\bar{d}}}\leftrightarrow a_{1(2)}^K. \nonumber
\end{equation}
The parameters $m_0^i$ $(i=\eta_{d\bar{d}(u\bar{u})}, \eta_{s\bar{s}},K)$
are defined as:
\begin{equation}
m_0^{\eta_{d\bar{d}(u\bar{u})}}\equiv m_0^\pi
\equiv \frac{m_{\pi}^2}{(m_u+m_d)}, \ \ \
m_0^{\eta_{s\bar{s}}}\equiv \frac{2M_K^2-m_{\pi}^2}{(2m_s)}, \ \ \ 
m_0^K\equiv \frac{M_K^2}{m_{d(u)}+m_s}. 
 \label{eq:19}
\end{equation}
Because there are large ambiguities in quark masses, parameters
defined in  Eq. (\ref{eq:19})
introduce considerable theoretical uncertainties. In our analysis, 
we use constraints for these parameters from analysis of other 
decay channels. 
In Ref.~\cite{LUY} branching ratios for $B \to \pi \pi$ 
are analysed in pQCD approach. The allowed region for 
$m_0^\pi$ is given as $1.1 \mbox{GeV} \leq 
m_0^\pi \leq 1.9 \mbox{GeV}$. 
Ref.~\cite{LKS}, which studies $B \to K \pi$ in pQCD,
gives the best fit value of $m_s=140$ MeV. 

We saw that the $SU(3)$ breaking effects were included 
through the decay constants and the wave functions in Eq. (\ref{eq:7-5}) and
Eq. (\ref{eq:19}), respectively.
In exact $SU(3)$ symmetry limit,
\begin{eqnarray}
f_{\pi}=f_x=f_y=f_K, \ \ \\
m_0^{\eta_{d\bar{d}(u\bar{u})}}=m_0^{s\bar{s}}=m_0^K,
\end{eqnarray}
Eq. (\ref{eq:3})is recovered.

Now we show our numerical results. 
The parameters which are used in our calculation are as follows: 
\begin{eqnarray}
&G_F=1.16639\times 10^{-5} \mbox{GeV}^{-2},
\Lambda_{\bar{MS}}^{(4)} =250  \ \mbox{MeV},
\alpha_s(M_Z)=0.117, \alpha_{em}=1/129, & \nonumber \\ 
&\tau_{B^0}=1.56 ps, \lambda=0.2196, A=0.819, R_b=\sqrt{\rho^2+\eta^2}=0.38,
\phi_3=90^{\circ}, & \nonumber \\
&M_W=80.2\ \mbox{GeV}, \ \ M_B=5.28  \ \mbox{GeV}, \ \ m_t=170  \ \mbox{GeV}, m_b=4.8  \ \mbox{GeV}, & \nonumber \\
&f_B=190  \ \mbox{MeV}, f_K=160  \ \mbox{MeV}, f_{\pi}=130  \ \mbox{MeV} & 
\nonumber
\end{eqnarray}
The B meson wave function is given as follows: 
\begin{eqnarray}
&\phi_B(x)=N_Bx^2(1-x)^2exp[-\frac{1}{2}(\frac{xM_B}{\omega_B})^2-
\frac{\omega_B^2b^2}{2}]& \nonumber \\
& N_B=91.7835  \ \mbox{GeV},& \nonumber 
\end{eqnarray}
where $\omega_B$ is a free parameter. 
At first, we show our result with 
the best fit parameter set which is obtained 
from analysis of $B \to \pi \pi$ and $B \to K \pi$ processes in pQCD 
approach ~\cite{LKS,LUY}: 
\begin{equation}
\{ m_0^{\eta_{d\bar{d}(u\bar{u})}}, m_s, \omega_B \}
=\{ 1.4\mbox{GeV}, 140\mbox{MeV}, 0.4 \mbox{GeV}\}
\end{equation}
Our result of the branching ratio for the above parameter set is 
\begin{eqnarray}
Br(\bar{B^0}\to \bar{K^0} \eta^{\prime})&=& 18 (16) \times 10^{-6}
\label{eq:21}\\
Br(\bar{B^0}\to \bar{K^0} \eta)&=& 0.44 (1.9) \times 10^{-6},
\label{eq:22}
\end{eqnarray}
for the pseudoscalar mixing angle $\theta_p=-20^{\circ}(-10^{\circ})$.
The computed amplitude for the each diagram in Fig. \ref{fig:5} is given
in Table \ref{table:1}. With the same set of parameters,
we obtain theoretical prediction for $\bar{B^0} \to \bar{K^0} \pi^0$ as
$Br(\bar{B^0} \to \bar{K^0} \pi^0)=9.5 \times 10^{-6}$.
\begin{table}
\begin{center}
\renewcommand{\arraystretch}{0.85}
\begin{tabular}{|c|l|l|c|l|} \cline{1-2}\cline{4-5}
$f_K\frac{X_{\eta^{\prime}}}{\sqrt{2}}F_e^{I}$ & -1.57 \ (-1.93)&  \ \ \ \ \ &
$f_K\frac{X_{\eta}}{\sqrt{2}}F_e^{I}$ & -2.26 \ (-1.95) \\
\cline{1-2}\cline{4-5}
$f_yY_{\eta^{\prime}}F_e^{II}$ & -4.03 \ (-3.50)  & \ \ \ \ \ &
$f_yY_{\eta}F_e^{II}$ & \ 2.81 \ (3.50)  \\
\cline{1-2}\cline{4-5}
$(f_xX_{\eta'}/\sqrt{2}\xi +
f_xX_{\eta'}/\sqrt{2} \atop +f_yY_{\eta'})F_e^{III}
$ & \ 2.16 \ \ (2.15) &  \ \ \ \ \ &
$(f_xX_{\eta}/\sqrt{2}\xi +
f_xX_{\eta}/\sqrt{2} \atop +f_yY_{\eta})F_e^{III} $ &  0.120 \ \ (-0.266)\\
\cline{1-2}\cline{4-5}
$\frac{V_u}{V_t}f_x\frac{X_{\eta^{\prime}}}{\sqrt{2}}F_e^{IV}$ & \ $i0.0334$ ($i0.0411$)   & \ \ \ \ \ &
$\frac{V_u}{V_t}\frac{X_{\eta}}{\sqrt{2}}F_e^{IV}$ & \ $i0.0481$ ($i0.0417$)   \\
\cline{1-2}\cline{4-5}
$f_B\frac{X_{\eta^{\prime}}}{\sqrt{2}}\mbox{Re}{F_a^{A}}$ &  \ 0.133 (0.163)   & \ \ \ \ \ &
$f_B\frac{X_{\eta}}{\sqrt{2}}\mbox{Re}{F_a^{A}}$ &  \ 0.191 (0.165)\\
\cline{1-2}\cline{4-5}
$f_B\frac{X_{\eta^{\prime}}}{\sqrt{2}}\mbox{Im}{F_a^{A}}$ & \ 0.734 \ \ (0.902)  & \ \ \ \ \ &
$f_B\frac{X_{\eta}}{\sqrt{2}}\mbox{Im}{F_a^{A}}$ & \ 1.06 \ \ (0.915)\\
\cline{1-2}\cline{4-5}
$f_BY_{\eta^{\prime}}\mbox{Re}{F_a^{B}}$ & \ 0.494 (0.428)   & \ \ \ \ \ &
$f_BY_{\eta}\mbox{Re}{F_a^{B}}$ &  -0.343 (-0.428)\\
\cline{1-2}\cline{4-5}
$f_BY_{\eta^{\prime}}\mbox{Im}{F_a^{B}}$ & \ 2.00 \ \ (1.74)   & \ \ \ \ \ &
$f_BY_{\eta}\mbox{Im}{F_a^{B}}$ &  -1.39 \ \ (-1.74) \\
\cline{1-2}\cline{4-5}
\end{tabular}
\end{center}
\caption[table1]{\vspace{-0.5cm}
The numerical result with the best fit parameter set,  \vspace{-0.5cm} 
$\{ m_0^{\eta_{d\bar{d}(u\bar{u})}}, m_s, \omega_B \}
=\{ 1.4\mbox{GeV}, 140\mbox{MeV}, 0.4\mbox{GeV} \}$
for the each term in Eq. (\ref{eq:20}). \vspace{-0.5cm} 
The pseudoscalar mixing angle is 
taken as $\theta_p=-20^{\circ}(-10^{\circ})$. 
The values are factored out by $10^{4}V_t$. }
\label{table:1}
\end{table}

Since we observe large imaginary part for annihilation diagrams we need to 
rewrite the $SU(3)$ relation 
in Eqs. (\ref{eq:12-5}) and (\ref{eq:13-5}) more precisely. 
Again ignoring the small amount of tree contribution, we obtain: 
\begin{eqnarray}
\mbox{Re}A(B^0\to K^0\eta)&=&-(X_\eta+r\sqrt{2}Y_\eta+r\sqrt{2}s)
\mbox{Re}A(B^0\to K^0\pi^0)   \label{eq:120} \\
\mbox{Re}A(B^0\to K^0\eta^{\prime})&=&
-(X_{\eta'}+r\sqrt{2}Y_{\eta'}+r\sqrt{2}s')
\mbox{Re}A(B^0\to K^0\pi^0)   \label{eq:130} \\
\mbox{Im}A(B^0\to K^0\eta)&=&-(X_\eta+r'\sqrt{2}Y_\eta)
\mbox{Im}A(B^0\to K^0\pi^0)   \label{eq:120-2} \\
\mbox{Im}A(B^0\to K^0\eta^{\prime})&=&-(X_{\eta'}+r'\sqrt{2}Y_{\eta'})
\mbox{Im}A(B^0\to K^0\pi^0)   \label{eq:130-2}
\end{eqnarray}
where 
\begin{eqnarray}
&r=\frac{f_yF_e^{II}+f_BReF_a^B}{f_KF_e^I+f_BReF_a^A}, 
\ \ \ \ \ \ 
r'=\frac{ImF_a^B}{ImF_a^A} & \\
&s'=\frac{(f_x\frac{X_{\eta^{\prime}}}{\sqrt{2}}\xi+
        f_x\frac{X_{\eta'}}{\sqrt{2}}+f_yY_{\eta'})F_e^{III}}
        {f_yF_e^{II}+f_BRe{F_a^B}}, \ \ \ \   
s=\frac{(f_x\frac{X_\eta}{\sqrt{2}}\xi+
        f_x\frac{X_{\eta}}{\sqrt{2}}+f_yY_{\eta})F_e^{III}}
        {f_yF_e^{II}+f_BRe{F_a^B}}& \label{eq:36-4}
\end{eqnarray}
Using the values listed in Table \ref{table:1}, the $SU(3)$ breaking 
effect $r^{(')}$, which is assumed as $r=1$ in Eq.(\ref{eq:3}),  
is calculated as $r=1.2$ and  $r'=1.3$ 
and the value of $s^{(')}$ which is proportional to the $SU(3)$ singlet 
contribution is calculated as  
      $s'=-0.50 (-0.50)$ and $s=-0.028 (0.059)$
for $\theta_p=-20^{\circ} (-10^{\circ})$. The electric penguin 
correction factor is obtained as $\xi=0.543$. 
To employ the new definition of the decay constant in $\eta-\eta^{\prime}$ 
system modifies the relation between $s$ and $s'$ as 
\begin{equation}
s'/s=\frac{(f_x\frac{X_{\eta^{\prime}}}{\sqrt{2}}\xi+
        f_x\frac{X_{\eta'}}{\sqrt{2}}+f_yY_{\eta'})}
{(f_x\frac{X_\eta}{\sqrt{2}}\xi+
        f_x\frac{X_{\eta}}{\sqrt{2}}+f_yY_{\eta})}, 
\end{equation}
which is assumed to be $s'/s=-\cos\theta_p/\sin\theta_p$ in Eq. (\ref{eq:3}).  
$s'$ always has minus sign. 
This is due to the sign difference between 
$F_e^{II}$ and $F_e^{III}$, which can be traced back to the relative size
of the  Wilson coefficients. 
This effect is also seen in calculations using vacuum saturation approximation 
~\cite{AliLuKr}. This fact implies that the $SU(3)$ singlet 
penguin contribution to $\bar{B^0} \to \bar{K^0} \eta^{\prime}$ tends
to decrease the branching ratio.

Before we show our numerical results for the different parameter sets 
of the wave functions, it might be convenient to summarize  
the trend of the size of the amplitudes for the variation of the parameters, 
diagram by diagram in  Fig. \ref{fig:5}. 
The variable parameters are $m^{\eta_{d\bar{d}(u\bar{u})}}_0$, 
$m^{\eta_{s\bar{s}}}_0$ and $m^K_0$ which depend on quark masses and 
$\omega_B$ which parameterizes momentum distribution of $b$ quark
in a B meson.  
The amplitudes $F_e^{I}$, Re$F_a^A$ and Im$F_a^A$ depend on 
parameters $m^{\eta_{d\bar{d}(u\bar{u})}}_0$ and $m^K_0$, 
$F_e^{II}$, Re$F_a^B$ and Im$F_a^B$ depend on 
parameters $m^{\eta_{s\bar{s}}}_0$ and $m^K_0$ and 
$F_e^{III}$ depends on $m^K_0$. All the amplitudes are increased 
when $\omega_B$ is decreased. Note that the branching ratio of 
$B^0 \to K^0 \pi^0$ are written in terms of $F_e^{I}$, 
Re$F_a^A$ and Im$F_a^A$. 

We first discuss the input parameter dependence of $s'$ and $s$. 
When we decrease $m_s$, which increases $m^{\eta_{s\bar{s}}}_0$ 
and $m^K_0$, the amplitudes $F_e^{II}$, 
Re$F_a^B$ and $F_e^{III}$ get enhanced simultaneously. 
When we decrease $\omega_B$, $F_e^{II}$ and $F_e^{III}$ increase while
Re$F_a^B$ remains unchanged.
Therefore, 
as we can see from Eq. (\ref{eq:36-4}), $s'$ and $s$ are quite
insensitive to the input parameters $m^{\eta_{s\bar{s}}}_0$, 
$m^K_0$ and $\omega_B$. 

Now let us see how we can obtain large 
$ Br(\bar{B^0}\to \bar{K^0} \eta^\prime)$.
To enhance $ Br(\bar{B^0}\to \bar{K^0} \eta^\prime)$, we need large $r$.
The value for $r$ is increased for smaller 
$m^{\eta_{d\bar{d}(u\bar{u})}}_0$, smaller $m_s$ or smaller $\omega_B$.  
First, let us try $\omega_B=0.3GeV$ which
is the lower limit considering other two body $B$ decays.
For example, we obtain
\begin{eqnarray}
 &r=1.2, \ \ \ r'=1.3, \ \ \ s=-0.032(0.054), \ \ \ s'=-0.49(-0.49) & 
\nonumber \\
 &Br(\bar{B^0}\to \bar{K^0} \pi^0)=21\times 10^{-6}, 
 \ \ \ Br(\bar{B^0}\to \bar{K^0} \eta^{\prime})=34(31)\times 10^{-6},& 
\nonumber
\end{eqnarray}
for $\theta_p=-20^{\circ} (-10^{\circ})$. 
In order to further increase $r$ and thus increase
$\bar{B^0}\to \bar{K^0} \eta^{\prime}$, we need to reduce 
$m_s$, however, it ends up also increasing 
$\bar{B^0}\to \bar{K^0} \pi^0$ branching ratio. This is unacceptable.
Therefore, we put back $\omega_B$ as 0.4GeV and try to enhance $r$ only by 
changing $m^{\eta_{d\bar{d}(u\bar{u})}}_0$ and $m_s$. 
In fact, the dependence of $r$ on $m^{\eta_{d\bar{d}(u\bar{u})}}_0$ is 
so weak that even if we take smaller value for 
$m^{\eta_{d\bar{d}(u\bar{u})}}_0$ the branching ratio for 
$\bar{B^0}\to\bar{K^0} \eta^{\prime}$ does not become large enough. 
So, the only possibility is to take a smaller value for $m_s$. 
The results evaluated with 
$\{m_s, \omega\} = \{100\mbox{MeV}, 0.4\mbox{GeV}\}$ are as follows: 
\begin{eqnarray}
&\mbox{For}\ m^{\eta_{d\bar{d}(u\bar{u})}}_0=1.9 \ \mbox{GeV}\hspace*{10cm}& 
\nonumber\\ 
&r=1.4, \ r'=1.3, \ s=-0.025(0.045),\ s'=-0.40(-0.40),& \label{eq:100} \\
&\mbox{For}\ m^{\eta_{d\bar{d}(u\bar{u})}}_0=1.4 \ \mbox{GeV}\hspace*{10cm}& \nonumber\\
&r=1.7, \ r'=1.5, \ s=-0.025(0.045), \ s'=-0.40(-0.40), \ & \label{eq:101}\\
&\mbox{For}\ m^{\eta_{d\bar{d}(u\bar{u})}}_0=1.1 \ \mbox{GeV}\hspace*{10cm}&\nonumber\\
&r=2.0, \ r'=1.6,\ s=-0.025(0.045), \ s'=-0.40(-0.40), \  & \label{eq:102} 
\end{eqnarray}
Obtained branching ratios for $\bar{B^0}\to \bar{K^0} \eta^{\prime}$ and 
$\bar{B^0}\to \bar{K^0} \pi^0$ are given in Table\ref{table:2}. 
Considering that BELLE reported a smaller branching ratio for 
$\bar{B^0}\to\bar{K^0} \eta^{\prime}$ and also that experimental value of 
$\bar{B^0}\to \bar{K^0} \pi^0$ has still large error and can be large, 
a situation such that $m^{\eta_{d\bar{d}(u\bar{u})}}_0=1.9 \ \mbox{GeV}$ 
can not be excluded.

For a comparison to other approaches, we give the obtained 
values of the form factors for different parameter; 
\begin{eqnarray}
F^{B\to\pi}_0(0)=(0.36, 0.30, 0.26) \ \ &:&
 \ m^{\eta_{d\bar{d}(u\bar{u})}}_0=(1.9, 1.4, 1.1)GeV \nonumber \\ 
F^{B\to K}_0(0)=(0.47, 0.35) \ \ &:&
 \ m_s=(100, 140)MeV  \nonumber 
\end{eqnarray}
where $\omega_b=0.4$GeV. 

In fact, there is another interesting aspect. Looking at Eq. (\ref{eq:12-5}), 
the first and the second term which are the dominant contributions 
have an opposite sign so that the relative variation of the 
branching ratio for $B^0 \to K^0 \eta$ is much larger than that for 
$B^0 \to K^0 \eta^{\prime}$ when $r$ varies. 
The branching ratio for $B^0 \to K^0 \eta$, which is considered to be 
negligible in Eq. (\ref{eq:3}), is enhanced greatly depending on the 
 parameters. We summarize our numerical 
results of the branching ratios for $\bar{B^0}\to \bar{K^0} \pi^0$, 
$\bar{B^0}\to\bar{K^0} \eta^{\prime}$ and $\bar{B^0}\to\bar{K^0} \eta$ 
in Table \ref{table:2}. 
We can see a large dependence of 
the branching ratio for $\bar{B^0}\to\bar{K^0} \eta$ on $\theta_p$. As 
is mentioned before, $\theta_p=-20^{\circ}$ is the smallest limit of 
the allowed region. Hence, our results for $\theta_p=-10^{\circ}$ 
indicate that $\bar{B^0}\to\bar{K^0} \eta$ process may be observed soon. 

\begin{table}
\renewcommand{\arraystretch}{0.8}
\begin{center}
\begin{tabular}{|c|c|c|c|}
\cline{1-4}
\ \ \ &$Br(\bar{B^0}\to\bar{K^0}\pi^0)$ &
$Br(\bar{B^0}\to\bar{K^0}\eta^{\prime})$ &
$Br(\bar{B^0}\to\bar{K^0}\eta)$       \\
\cline{1-4}
$m^{\eta_{d\bar{d}(u\bar{u})}}_0=1.9 \ \mbox{GeV}$  &
$21\times 10^{-6}$&$50(45)\times 10^{-6}$&$2.4(7.5)\times 10^{-6}$   \\
\cline{1-4}
$m^{\eta_{d\bar{d}(u\bar{u})}}_0=1.4 \ \mbox{GeV}$  &
$15\times 10^{-6}$&$45(39)\times 10^{-6}$&$4.6(11)\times 10^{-6}$ \\
\cline{1-4}
$m^{\eta_{d\bar{d}(u\bar{u})}}_0=1.1 \ \mbox{GeV}$  &
$11\times 10^{-6}$&$41(35)\times 10^{-6}$&$7.0(13)\times 10^{-6}$ \\
\cline{1-4}
\end{tabular}
\end{center}
\caption[table2]{\vspace{-0.5cm}
The numerical results for the branching ratios of 
$\bar{B^0}\to \bar{K^0} \pi^0$, \vspace{-0.5cm}
$\bar{B^0}\to\bar{K^0} \eta^{\prime}$ and $\bar{B^0}\to\bar{K^0} \eta$ 
with parameters tuned to increase $\bar{B^0}\to\bar{K^0} \eta^{\prime}$.
\vspace{-0.5cm}$\{m_s, \omega\} = \{100\mbox{MeV}, 0.4\mbox{GeV}\}$ 
and different $m^{\eta_{d\bar{d}(u\bar{u})}}_0$. 
These parameter sets \vspace{-0.5cm} are allowed by the 
pQCD analysis for $B \to \pi \pi$ 
and $B \to K \pi$ processes. 
The pseudoscalar mixing angle is taken as
$\theta_p=-20^{\circ} (-10^{\circ})$. }
\label{table:2}
\end{table}

In conclusion, we examined the large branching ratio of 
$B \to K \eta^{\prime}$ process using  $SU(3)$ relation in general form, 
Eqs. (\ref{eq:12-5}) and (\ref{eq:13-5}). If there is a large 
$SU(3)$ breaking effect, which means that $r$ is much larger than 1, or 
there is a large $SU(3)$ singlet penguin contribution, which means $s'$ and 
$s$ are very large, Eqs. (\ref{eq:12-5}) and (\ref{eq:13-5}) imply 
that we would have large branching ratio for 
$\bar{B^0}\to\bar{K^0} \eta^{\prime}$. 
We computed $r$, $s'$ and $s$ as well as branching ratios for 
$\bar{B^0}\to\bar{K^0} \pi^0$, $\bar{B^0}\to\bar{K^0} \eta^{\prime}$ and 
$\bar{B^0}\to\bar{K^0} \eta$ processes in pQCD approach. 
$s'$ is found to contribute destructively to 
the other dominant contributions to $\bar{B^0}\to\bar{K^0} \eta^{\prime}$ 
process. Our numerical result in Table \ref{table:2} 
indicates that in a case that the experimental data for 
$\bar{B^0}\to\bar{K^0} \eta^{\prime}$ will come close to the lower limit of 
BELLE data, $\bar{B^0}\to\bar{K^0} \eta^{\prime}$ problem can be 
understood in the standard model. However, in this case, 
the correlation of the experimental data for 
$\bar{B^0}\to\bar{K^0} \eta^{\prime}$ to the experimental data for 
$\bar{B^0}\to\bar{K^0} \pi^0$ and $\bar{B^0}\to\bar{K^0} \eta$  
has to be examined carefully. 
In particular, considering that the relatively large value of 
$\theta_p$ which is close to $-10^{\circ}$ is favored by recent experiments, 
we have to keep in mind that $Br(\bar{B^0}\to\bar{K^0} \eta)$ may not
be so small.
If $Br(\bar{B^0}\to\bar{K^0} \eta^{\prime})$ remains high 
at its present value or the combination of the experimental data for 
$\bar{B^0}\to\bar{K^0} \pi^0$, 
$\bar{B^0}\to\bar{K^0} \eta^{\prime}$ and $\bar{B^0}\to\bar{K^0} \eta$  
deviate from our result in Table \ref{table:2}, 
it may imply that we need modify our understanding of 
$\eta^{\prime}$. 

\begin{figure}[ht]
 \begin{center}
\begin{center} \scalebox{0.7}{\begin{picture}(400,100)(0,0)
\ArrowLine(15,10)(165,10)
\ArrowLine(15,90)(165,90)
\Photon(75,90)(75,55){4}{3}
\ArrowLine(75,55)(165,65)
\ArrowLine(75,55)(165,35)
\put(5,50){$B$}
\put(45,93){$b$\ \ \ \ \ \ \ \ \ \ \ \ \ \ \ \ \ \ \ \ \ $c$}
\put(160,70){$c$} \put(160,25){$s$} \put(60,70){$W$}
\put(45,0){$d$\ \ \ \ \ \ \ \ \ \ \ \ \ \ \ \ \ \ \ \ \ $d$}
\put(170,80){$\eta_c\Longrightarrow\eta^{\prime}$} \put(170,15){$K$}
\ArrowLine(250,10)(400,10)
\ArrowLine(250,60)(400,60)
\GlueArc(367,60)(30,90,180){5}{6}
\GlueArc(367,10)(100,90,180){5}{12}
\ArrowLine(367,110)(400,105) \Line(367,110)(367,90)
\ArrowLine(367,90)(400,95)
\put(240,35){$B$} \put(400,35){$K$} \put(405,98){$\eta^{\prime}$}
\put(265,63){$b$\ \ \ \ \ \ \ \ \ \ \
\ \ \ \ \ \ \ \ \ \ \ \ \ \ \ \ \ \ \ \ \ $s$}
\put(275,0){$d$\ \ \ \ \ \ \ \ \ \ \ \ \ \ \ \ \ \ \ \ \ $d$}
\put(310,80){$g$} \put(365,75){$g$}
\end{picture}} \end{center}

 \end{center}
\vspace{-1cm}%)
 \caption[fig1]
{\vspace{-0.5cm}
(a) Intrinsic charm contribution,
(b) $SU(3)$ singlet contribution}
 \label{fig:1}
\end{figure}
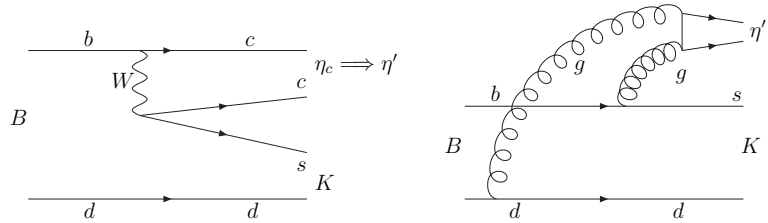

Finally, we would like to make a comment on
two suggested mechanisms to explain
the large branching ratio for $B \to K \eta^{\prime}$,
\begin{description}
\item[(1)]
Intrinsic charm contribution: \\
The Cabbibo allowed $b\to uc\bar{c}$ process can contribute to
$B \to K \eta^{\prime}$ if there is intrinsic $c\bar{c}$ content
in $\eta^{\prime}$ (see Fig. \ref{fig:1}(a))~\cite{HZ,PET2}.
\end{description}
The amount of the
$c\bar{c}$ content in $\eta^{\prime}$ which is parameterized by
decay constant $f_c^{\eta^{\prime}}$ as:
\begin{equation}
if_c^{\eta^{\prime}}p_{\mu} = \langle 0| \bar{c}\gamma_{\mu}
\gamma_5 c|\eta^{\prime}(p) \rangle
\end{equation}
Decay constant $f_c^{\eta^{\prime}}$ is obtained
from the radiative $J/\psi$ decay and two photon process of $\eta^{\prime}$
. It was found that the numerical result for the
intrinsic charm contribution to $B \to K \eta^{\prime}$ decay was very
small~\cite{KF,KFS}.
\begin{description}
\item[(2)]
The $SU(3)$ singlet contributions: \\
$B\to K \eta^{\prime}$ is produced by fusion of gluons,
one gluon from $b\to sg$ process and another one from spectator.
(see Fig. \ref{fig:1}(b))
~\cite{SONI,AKS,AK}.
\end{description}
According to the paper ~\cite{KOU}, there is still a possibility
that $\eta^{\prime}$ includes at most 26 \% of pure gluonic state,
gluonium. The contribution of the diagram in which
two gluons in Fig. \ref{fig:1}(b) are directly
attached to gluonium in $\eta^{\prime}$ instead of attached to
triangle quark loop may be important for $B \to K \eta^{\prime}$ decay.

\newpage
\noindent
{\bf Note added in proof}\\
After this paper was submitted for publication, BELLE 
and BABAR announced new data: 

\vspace{0.5cm}
\hspace{-3cm}
\begin{minipage}{20cm}
\begin{center}
\begin{tabular}{|c|c|c|}
\cline{1-3}
\ \ \ &$Br(B^0\to K^0\eta^{\prime})$ &
$Br(B^+\to K^+\eta^{\prime})$    \\
\cline{1-3}
BELLE ~\cite{BELLE2001} &
$(55^{+19}_{-16} \pm 8) \times 10^{-6}$& 
$(79^{+12}_{-11} \pm 9) \times 10^{-6}$ \\ 
\cline{1-3}
BABAR ~\cite{BABAR2001} &
$(42^{+13}_{-11}\pm 4)\times 10^{-6}$&$(70\pm 8\pm 5)\times 10^{-6}$  \\
\cline{1-3}
\end{tabular}
\end{center}
\end{minipage}

\vspace{1.5cm} 
{\bf \large Acknowledgments}\\
The work was supported in part by Grant-in Aid for Special Project
Research (Physics of CP Violation), by Grant-in Aid for Scientific
Research from the Ministry of Education, Science and Culture of Japan.
We acknowledge pQCD group members:
Y.Y. Keum, T.Kurimoto, H-n. Li, C.D. L\"{u},
K. Ukai, N. Shinha, R. Sinha M.Z. Yang and T. Yoshikawa
for fruitful discussions. E.K. also thanks A. Sugamoto and Y.D. Yang for
interesting discussions.

%%%%%%%%%%%%%%%% References %%%%%%%%%%%%%%%%%%%%%%%

%\newpage

\vfill\eject

\end{document}